\documentclass[lettersize,journal]{IEEEtran}
\IEEEoverridecommandlockouts
\usepackage[utf8]{inputenc}
\usepackage{multirow}
\usepackage{array}
\usepackage{booktabs}
\usepackage{graphicx}
\usepackage{balance}
\usepackage[hidelinks]{hyperref}
\usepackage{comment}
\usepackage{enumitem}
\usepackage{ragged2e}
\usepackage{url}
\usepackage{stfloats}
\usepackage{tikz}
\usepackage{amsmath,amssymb}
\usepackage{listings}
\usepackage{xcolor}
\usepackage{colortbl}
\usepackage[nobreak,space,compress]{cite}

\def\BibTeX{{\rm B\kern-.05em{\sc i\kern-.025em b}\kern-.08em T\kern-.1667em\lower.7ex\hbox{E}\kern-.125emX}}

\usetikzlibrary{shapes.geometric, positioning, arrows.meta, calc, backgrounds, shadows}
\newcommand{\dataset}{\textsc{Doc2CI}}
\newcommand{\kappaverdict}{0.78}
\newcommand{\kappabin}{0.81}
\newcommand{\relsub}{77}

\newcommand{\ftbase}{Qwen2.5-Coder-7B-Instruct}
\newcommand{\ftbaseL}{Qwen2.5-Coder-32B-Instruct}
\newcommand{\ftmodel}{\textsc{Doc2CI\mbox{--}FT}}
\newcommand{\srmodel}{\textsc{Doc2CI\mbox{--}SR}}
\newcommand{\ftsrmodel}{\textsc{Doc2CI\mbox{--}FT\mbox{+}SR}}

\IfFileExists{fontawesome5.sty}{
  \usepackage{fontawesome5}
  \newcommand{\mcode}{\faCode}
  \newcommand{\mgeneral}{\faCommentDots}
}{
  \newcommand{\mcode}{\textsf{[code]}}
  \newcommand{\mgeneral}{\textsf{[gen]}}
}

\definecolor{DarkGreen}{RGB}{1,150,32}
\lstdefinestyle{PromptStyle}{
  basicstyle=\ttfamily\scriptsize, breaklines=true, frame=single, captionpos=b,
  rulecolor=\color{gray}, escapeinside={(*@}{@*)},
  moredelim=**[is][\color{DarkGreen}\bfseries]{@d@}{@},
  moredelim=**[is][\color{blue}\bfseries]{@u@}{@},
  moredelim=**[is][\color{black}\bfseries]{@s@}{@}
}

\usepackage{tcolorbox}
\tcbset{rqbox/.style={colback=black!5!white, boxrule=0pt, colframe=black!5!white,
    sharp corners, boxsep=5pt, left=4pt, right=4pt, top=3pt, bottom=3pt, fontupper=\normalsize}}

\begin{document}

\title{Doc2CI: A Multi-Service Study of CI Configuration Generation Using Large Language Models}

\author{
  Taher A. Ghaleb~\href{https://orcid.org/0000-0001-9336-7298}{\includegraphics[scale=0.06]{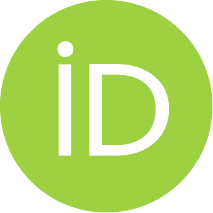}}

  \thanks{Taher A. Ghaleb is with the Department of Computer Science, Trent University, Peterborough, Ontario, Canada (e-mail: taherghaleb@trentu.ca).}
}

\markboth{}
{Ghaleb: Doc2CI: A Multi-Service Study of CI Configuration Generation Using Large Language Models}

\maketitle

\begin{abstract}
Adopting Continuous Integration (CI) often requires writing YAML configurations that are error-prone and challenging to maintain. Despite increasing LLM use in software engineering, their ability to generate CI configurations from natural language across services and model families remains unclear.
This paper presents a large empirical study on using LLMs to generate CI configurations. We introduce \dataset{}, a benchmark of $3{,}363$ description-to-YAML pairs collected from the official documentation of four CI services, and evaluate $14$ open-weight models from $7$B--$34$B parameters together with GPT-4o and GPT-4.1, producing over $53{,}000$ configurations. We assess both reference alignment and schema validity to determine whether the generated configurations are structurally valid. We further develop a failure taxonomy from a manual analysis of $385$ configurations and examine why LLMs disagree.
Across models and services, exact reference reproduction never exceeds $3.1\%$, and while $97\%$ of outputs parse as YAML, only $71\%$ satisfy service schemas. Larger models improve structural validity, but code specialization provides no consistent advantage over comparable general models. Model differences are driven largely by output completeness: for the same request, some models generate the expected fragment while others produce a full workflow. Finally, a training-free schema-guided repair method improves schema validity to $94\%$, while fine-tuning improves similarity to documentation but reduces standalone validity. This suggests that similarity and validity are distinct objectives for CI generation and motivate schema-aware evaluation and tooling for LLM-based configuration generation.
\end{abstract}


\section{Introduction}
\label{sec:intro}

\IEEEPARstart{C}{ontinuous} Integration (CI) is central to modern software development, automating building and testing of source code on every change~\cite{Fowler_CI}. Configuring CI requires developers to write YAML files that are challenging~\cite{saroar2023developers} both syntactically, due to strict indentation and structure, and semantically, due to service-specific keywords, job matrices, reusable components, and third-party integrations~\cite{ghactions_workflow_syntax}. This is particularly apparent for developers who are new to CI concepts or a certain service, and poor configurations can be associated with long build durations and failures~\cite{ghaleb2019duration,ghaleb2022interplay}.

Large Language Models (LLMs) now automate many software engineering tasks, such as code generation, documentation, and bug fixing. However, generating CI configurations remains challenging because CI files are non-executable, declarative YAML with domain-specific semantics rather than imperative code~\cite{pujar2023automated}, so success in code generation does not necessarily generalize to this setting. Prior work studied CI configuration challenges~\cite{hilton2016usage,widder2019conceptual,zampetti2020empirical,vassallo2020configuration} and explored automated generation in limited contexts: Mastropaolo et al.~\cite{mastropaolo2024toward} complete partial GitHub Actions workflows, while Zhang et al.~\cite{zhang2024effectiveness} evaluate LLMs on manually crafted, syntax-focused scenarios. A prior single-service evaluation~\cite{ghaleb2025canllmsci} showed that LLMs generate GitHub Actions configurations with at most $3\%$ exact reference reproduction and that models pretrained on code do not outperform comparable general models. However, it remains unknown whether these findings generalize across CI services, model scales, and model families, or why models fail.

This paper addresses these gaps through a large empirical study of LLM-powered CI configuration generation. We construct \dataset{}, a benchmark of $3{,}363$ \emph{description-to-YAML} pairs curated from the official documentation of GitHub Actions, CircleCI, GitLab CI, and Travis CI. We evaluate $14$ open-weight models from $7$B--$34$B parameters, pairing general families with code-specialized counterparts, together with GPT-4o and GPT-4.1, under zero-shot prompting~\cite{kojima2022large}, producing $53{,}808$ configurations. Unlike prior evaluations focused primarily on similarity, we assess both reference alignment through eight lexical and structural metrics and validity through YAML parsing and conformance to service schemas. We further manually analyze a stratified sample of $385$ configurations using open coding~\cite{strauss1990basics} to derive a failure taxonomy and examine the causes of model disagreement.

Our results show that exact reproduction rarely exceeds $3\%$: although $97\%$ of configurations parse, only $71\%$ satisfy service schemas. Larger models improve structural accuracy, but similarity gains are limited, and code specialization offers no overall advantage over comparable general models. Performance varies substantially by the target CI service, with CircleCI being significantly harder than the others. Failures are driven mainly by misinterpretation and omission, while model disagreement primarily reflects differences in output completeness, from minimal fragments to complete workflows. Finally, schema-guided repair raises validity to $94\%$, whereas fine-tuning improves similarity at the cost of standalone validity, showing that these objectives must be optimized separately.

\medskip\noindent\textbf{Contributions.} This paper makes the following contributions.
\begin{itemize}[leftmargin=12pt, itemsep=1pt]
    \item We introduce \dataset{}, a benchmark of $3{,}363$ description-to-YAML pairs across four CI services for evaluating natural language-based CI configuration generation.
    \item We conduct a large-scale evaluation of $16$ LLMs and $53{,}808$ configurations, showing that high reference similarity does not necessarily imply valid CI configurations.
    \item We study model scale and code specialization, showing that scale improves structural validity while code specialization provides no aggregate advantage and yields benefits only at larger scales within a model family.
    \item We develop an LLM failure taxonomy through open coding of $385$ configurations, revealing that failures are dominated by misinterpretation and omission, while LLM disagreement arises from output completeness.
    \item We propose two complementary approaches for CI generation through fine-tuning (which improves reference similarity) and schema-guided repair (which improves validity), with their combination achieving the best trade-off.
\end{itemize}

\noindent\textbf{Paper organization.}
Section~\ref{sec:background} gives background. Section~\ref{sec:design} describes our study design. Section~\ref{sec:results} reports results across six research questions. Section~\ref{sec:discussion} discusses implications and validity threats. Section~\ref{sec:related} presents related work. Finally, Section~\ref{sec:conclusion} concludes the paper and outlines future work.

\section{Background}
\label{sec:background}

\noindent\textbf{Continuous Integration and its configurations.}
CI merges developer changes frequently into a shared repository, triggering automated builds and tests for fast feedback~\cite{Fowler_CI}. These workflows are encoded in YAML, but each service defines its own vocabulary and structure: GitHub Actions uses \texttt{on} triggers and \texttt{jobs} with \texttt{steps} that call reusable \texttt{actions}; CircleCI uses \texttt{jobs}, \texttt{workflows}, \texttt{orbs}, and \texttt{executors}; GitLab CI uses top-level job names with \texttt{stages}, \texttt{rules}, and \texttt{include}; and Travis CI uses \texttt{language}, \texttt{script}, and \texttt{deploy} providers. The same intent can demand very different configurations across services. For example, a developer migrating a mature project from Travis CI to GitHub Actions may use an LLM to generate a workflow from a brief build description. The file may look fluent and on-topic yet still leak Travis-specific idioms into GitHub Actions, omit a required keyword, or produce only a fragment rather than a runnable configuration. These failures are not captured by similarity measures alone, as they appear only through parsing, schema validation, or expert review, which novice developers typically lack. This gap between fluent generation and reliable configuration motivates our study.

\smallskip\noindent\textbf{Large Language Models for configuration.}
Transformer-based LLMs such as the GPT, Llama, Gemma, and Qwen families are trained on large text and code corpora and have been applied to generating configuration and infrastructure files from descriptions~\cite{rosa2023automatically,mehta2023automated}. Two factors structure the model landscape we study: \emph{scale}, the parameter count, and \emph{specialization}, whether a model is general or pretrained on source code. A third, often conflated with specialization, is \emph{instruction tuning}, whether a model is trained to follow instructions rather than merely continue text. Distinguishing these factors is central to our analysis.

\smallskip\noindent\textbf{Evaluating generated YAML.}
Given that CI YAML is non-executable and sensitive to structure, no single metric captures quality. Lexical metrics such as BLEU~\cite{papineni2002bleu}, CrystalBLEU~\cite{eghbali2022crystalbleu}, ROUGE-L~\cite{lin2004rouge}, and chrF~\cite{popovic2015chrf} capture token and character overlap. Embedding-based cosine similarity~\cite{reimers2019sentence} captures semantic closeness, while structural metrics such as tree edit distance~\cite{zhang1989simple} and canonicalized Levenshtein distance capture layout. Validity is orthogonal: a configuration may resemble a reference yet fail to parse, or parse yet violate the service schema.

\section{Study Design}
\label{sec:design}
\begin{figure*}[t]
\centering
\vspace{-7pt}
\resizebox{\textwidth}{!}{
\begin{tikzpicture}[font=\footnotesize,
  cyl/.style={cylinder, shape border rotate=90, aspect=0.22, draw,
              minimum width=22mm, minimum height=13mm, align=center, inner sep=2pt},
  box/.style={rectangle, rounded corners=2pt, draw, align=center,
              minimum height=9mm, text width=21mm, inner sep=3pt},
  c-amber/.style ={fill={rgb,255:red,250;green,238;blue,218},
                 draw={rgb,255:red,133;green,79;blue,11},
                 text={rgb,255:red,65;green,36;blue,2}},
  c-gray/.style  ={fill={rgb,255:red,241;green,239;blue,232},
                 draw={rgb,255:red,95;green,94;blue,90},
                 text={rgb,255:red,44;green,44;blue,42}},
  c-blue/.style  ={fill={rgb,255:red,230;green,241;blue,251},
                 draw={rgb,255:red,24;green,95;blue,165},
                 text={rgb,255:red,4;green,44;blue,83}},
  c-purple/.style={fill={rgb,255:red,238;green,237;blue,254},
                 draw={rgb,255:red,83;green,74;blue,183},
                 text={rgb,255:red,38;green,33;blue,92}},
  c-teal/.style  ={fill={rgb,255:red,225;green,245;blue,238},
                 draw={rgb,255:red,15;green,110;blue,86},
                 text={rgb,255:red,4;green,52;blue,44}},
  stackblue/.style={box, c-blue, text width=44mm, align=left, minimum height=14mm,
                 double copy shadow={shadow xshift=2.4pt, shadow yshift=2.4pt,
                   opacity=1, fill={rgb,255:red,230;green,241;blue,251},
                   draw={rgb,255:red,24;green,95;blue,165}}},
  stackteal/.style={box, c-teal, text width=44mm, align=left, minimum height=14mm,
                 double copy shadow={shadow xshift=2.4pt, shadow yshift=2.4pt,
                   opacity=1, fill={rgb,255:red,225;green,245;blue,238},
                   draw={rgb,255:red,15;green,110;blue,86}}},
  stackpurple/.style={box, c-purple, text width=44mm, align=left, minimum height=14mm,
                 double copy shadow={shadow xshift=2.4pt, shadow yshift=2.4pt,
                   opacity=1, fill={rgb,255:red,238;green,237;blue,254},
                   draw={rgb,255:red,83;green,74;blue,183}}},
  ar/.style={-{Latex[length=2mm]}, thin,
             color={rgb,255:red,136;green,135;blue,128}},
  ttl/.style={font=\small\bfseries\itshape,
              text={rgb,255:red,95;green,94;blue,90}}]
  \node[cyl, c-amber]  (docs)   at (0,   0)  {\textbf{CI docs:}\\GitHub Actions\\CircleCI\\GitLab CI\\Travis CI};
  \node[box, c-gray]   (curate) at (3.1, 0)  {Curate\\pairs};
  \node[cyl, c-teal]   (bench)  at (6.1, 0)  {\dataset{}\\3,363 pairs};
  \node[box, c-purple, text width=27mm] (gen) at (9.8, 0)
        {\textbf{Prompt 16 LLMs:}\\2 proprietary,\\14 open-weight\\(7 general, 7 code;\\small/medium/large)};
  \node[cyl, c-blue]   (out)    at (13.6, 0) {53,808\\configurations};

  \node[box, c-blue, text width=30mm] (sim) at (17.6, 2)
        {\textbf{Similarity:}\\Cosine, Euclidean,\\BLEU, CrystalBLEU,\\ROUGE-L, chrF,\\Tree-Edit, Levenshtein};
  \node[box, c-blue, text width=30mm] (val) at (17.6, 0.4)
        {\textbf{Validity:} Parse, Schema};
  \node[box, c-teal, text width=30mm] (sample) at (17.6, -1.2)
        {\textbf{Open coding:}\\Stratified samples};
  \node[box, c-purple, text width=30mm] (ft) at (17.6, -3.1)
        {\textbf{Fine-tune \& Schema repair}\\(\dataset{} train/test)};

  \draw[ar] (docs)--(curate);   \draw[ar] (curate)--(bench);   \draw[ar] (bench)--(gen);
  \draw[ar] (gen)--(out);
  \draw[ar] (out.east) to[bend left=16]  (sim.west);
  \draw[ar] (out.east) to[bend left=4]   (val.west);
  \draw[ar] (out.east) to[bend right=14] (sample.west);
  \draw[ar] (bench.south) |- (ft.west);

  \node[stackblue] (rqA) at (23.9, 1.45)
     {\textbf{RQ1} Generation quality\\
      \textbf{RQ2} Scale and specialization\\
      \textbf{RQ3} Target CI service};
  \node[stackteal] (rqB) at (23.9, -1.2)
     {\textbf{RQ4} Failure taxonomy\\
      \textbf{RQ5} Models disagreement};
  \node[stackpurple] (rqC) at (23.9, -3.09)
     {\textbf{RQ6} Fine-tuning and Schema-guided repair};

  \draw[ar] (sim.east)    to[bend left=8]  (rqA.west);
  \draw[ar] (val.east)    to[bend right=8] (rqA.west);
  \draw[ar] (sample.east) -- (rqB.west);
  \draw[ar] (ft.east)     -- (rqC.west);

  \draw[dashed, color={rgb,255:red,180;green,178;blue,170}] (7.7,  3.3) -- (7.7,  -4);
  \draw[dashed, color={rgb,255:red,180;green,178;blue,170}] (15.4, 3.3) -- (15.4, -4);
  \draw[dashed, color={rgb,255:red,180;green,178;blue,170}] (21.0, 3.3) -- (21.0, -4);
  \node[ttl] at (3.4,  3.0) {Curate \dataset{}};
  \node[ttl] at (11.7, 3.0) {Generate};
  \node[ttl] at (17.6, 3.4) {Score and adapt};
  \node[ttl] at (23.9, 3.0) {Research questions};
\end{tikzpicture}}
\vspace{-15pt}
\caption{Overview of our study. Cylinders denote datasets, boxes denote processing steps, and stacked shapes group research questions. We curate description-to-YAML pairs from four CI services into \dataset{}, generate configurations with $16$ models, and analyze outputs through quantitative evaluation (RQ1--RQ3), open coding of stratified samples (RQ4--RQ5), and fine-tuning with schema-guided repair (RQ6).}
\vspace{-7pt}
\label{fig:overview}
\end{figure*}
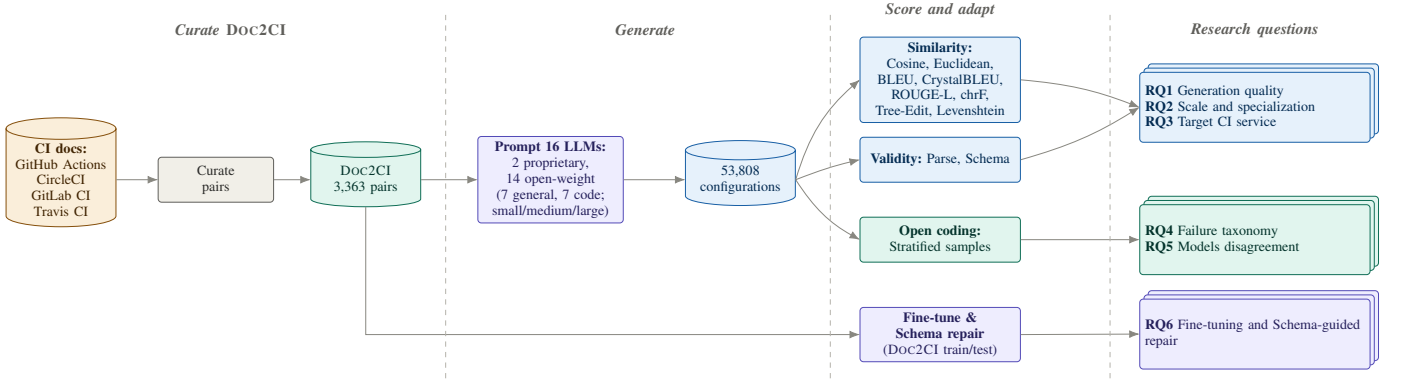

Fig.~\ref{fig:overview} outlines our study: we curate description-to-YAML pairs from four CI services, prompt $16$ models to generate a configuration for each description, and score each output for similarity and validity. Three analyses address our six research questions: quantitative scores address RQ1 to RQ3, open coding of a stratified sample addresses RQ4 and RQ5, and fine-tuning with schema-guided repair addresses RQ6. All data, generated outputs, scripts, and detailed results are in our replication package~\cite{our_replication_package}.
Each generation, fine-tuning, repair, and similarity analysis runs on a dedicated compute node in a shared cluster, each with one NVIDIA H100 GPU (80\,GB HBM3), an AMD EPYC 9654 CPU (96 cores), $755$\,GB RAM, Rocky Linux 9.8, PyTorch 2.9.0, and CUDA 12.6.

\subsection{The \dataset{} Benchmark}
\label{sec:data}

\begin{figure}[ht]
\centering
\fbox{\includegraphics[width=\linewidth]{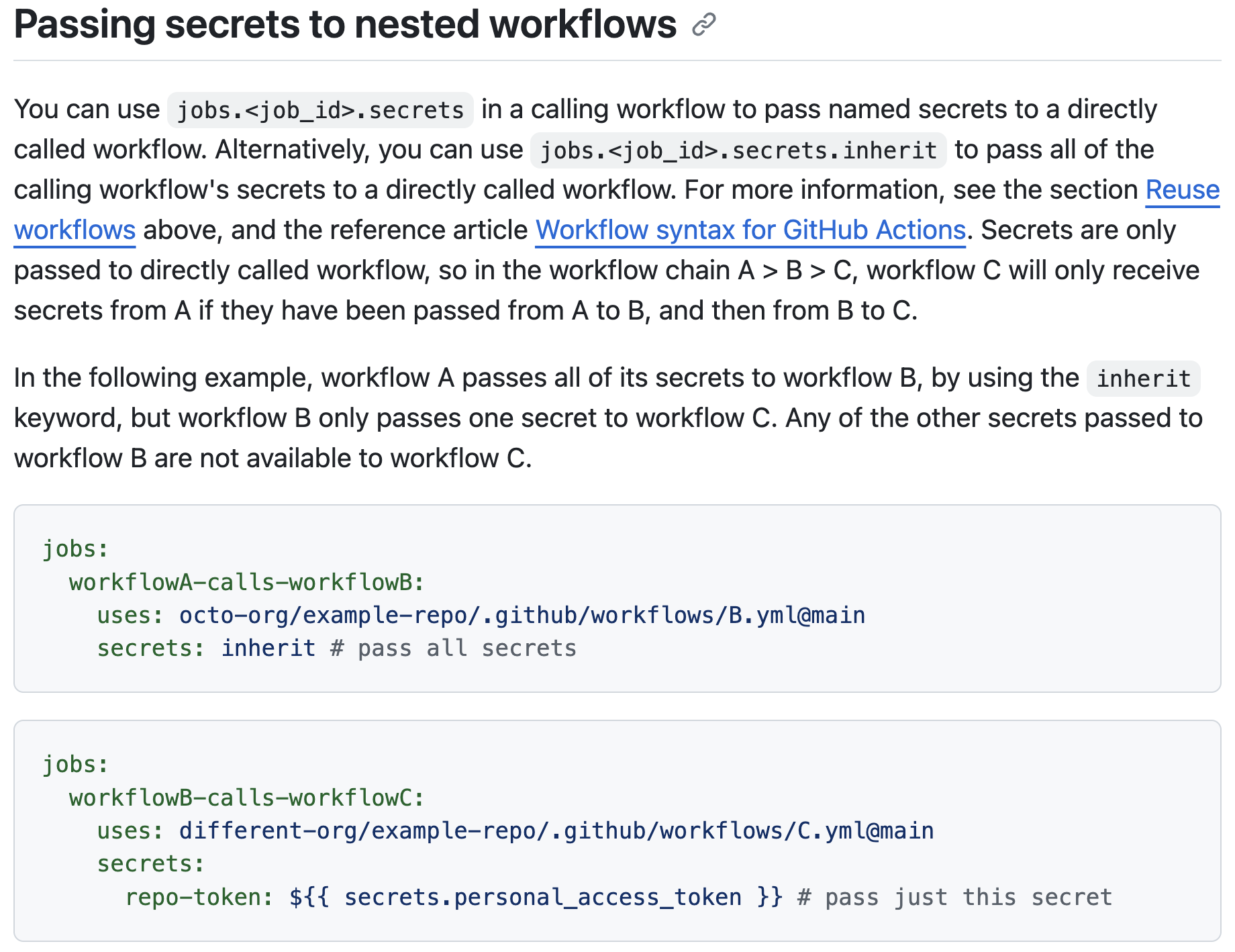}}
\caption{Example of a \dataset{} entry from the GitHub Actions documentation. The section title and explanatory prose form the \emph{description}, while each YAML block is a \emph{reference} configuration. Here, one \emph{description} maps to two valid \emph{references}: one inherits all secrets, and one forwards a single secret.}
\vspace{-7pt}
\label{fig:example}
\end{figure}

We developed a Selenium-based crawler to traverse the official documentation of GitHub Actions, CircleCI, GitLab CI, and Travis CI, following internal links and extracting instructional content while discarding navigation, sidebars, and footers. These docs explain keywords step by step and include example configurations and best practices, making them a natural source of \emph{description-to-YAML} pairs. Each pair consists of a task-oriented natural language \emph{description} (the section title plus its explanatory text) and a \emph{reference YAML} that implements the described behavior. We manually validated the pairs, fixed indentation and whitespace, and removed entries with empty descriptions.
Yet, the same information can appear multiple times in documentation, but for different reasons that require different treatments. First, providers reorganize and heavily cross-link their docs, so an identical section, with the same description and reference, is therefore reachable via several URLs, and the crawler collects each copy. We remove these exact-duplicate pairs, which account for $974$ of the initial $4{,}337$ crawled pairs ($22.5\%$). Second, a section can document a single task using multiple example configurations, and the same description can legitimately map to different reference snippets, as illustrated in Fig.~\ref{fig:example}, where one description is accompanied by a configuration that inherits all secrets, and another that forwards only one. We keep these as distinct pairs, since each generated output is scored against its own reference, and merge them into a single target only when fine-tuning (Section~\ref{sec:rq6}). After removing exact duplicates, \dataset{} contains $3{,}363$ unique pairs ($3{,}317$ unique descriptions): $725$ for GitHub Actions, $786$ for CircleCI, $841$ for GitLab CI, and $1{,}011$ for Travis CI. Descriptions and reference YAML configurations vary in complexity, from a single keyword to multi-job workflows with matrices and conditional execution.

One characteristic of documentation-derived references shapes our interpretation throughout: many references are minimal fragments that illustrate one keyword (e.g., a single-line \texttt{cache: pip} or a two-line \texttt{permissions} block), and some are non-workflow artifacts (an \texttt{action.yml} metadata file, a Helm \texttt{values.yaml}, or a Kubernetes manifest) or, in migration guides, snippets in another tool's language.

\subsection{Selected Models}
\label{sec:models}

We retain the six models evaluated in the prior single-service study~\cite{ghaleb2025canllmsci}: two proprietary models (GPT-4o~\cite{openai2023gpt4o} and GPT-4.1~\cite{openai2025gpt41}) and two general--code model pairs (Llama3.1-8B~\cite{touvron2023llama} with CodeLlama-7B~\cite{roziere2023codellama}, and Gemma3-12B~\cite{anil2024gemma} with CodeGemma-7B~\cite{zhao2024codegemma}). This allows direct comparison with the earlier results while expanding the evaluation to parameter scale and code specialization. To vary each factor cleanly, we add the Qwen2.5 family, the one open family with size-matched general and code-specialized checkpoints at $7$B, $14$B, and $32$B, yielding three directly comparable general-versus-code pairs with family and scale fixed and specialization varying. Around this basis, we include all other released checkpoints from the selected model families within a locally runnable $7$B--$34$B range: CodeLlama at $13$B and $34$B, Gemma3 at $27$B, and Gemma2 at $9$B. This yields $14$ open-weight models grouped into three size tiers: small ($7$--$8$B), medium ($9$--$15$B), and large ($>15$B). The tier boundaries follow the available model sizes, with no checkpoint between $14$B and $27$B and a natural separation between the $7$--$8$B and $9$--$14$B clusters.

This setting is not a complete family-by-tier grid, as released checkpoints do not cover all combinations. In the $7$--$34$B general range, only Llama3.1-$8$B is available (the next checkpoint is $70$B), and no single Gemma generation spans all tiers, requiring Gemma2-$9$B and Gemma3-$12$B/$27$B. CodeGemma is available only at $7$B. Consequently, only the three Qwen2.5 pairs match on both scale and generation; Llama and Gemma comparisons also vary in size or generation. We therefore use Qwen2.5 for the controlled specialization test and treat the remaining families as broader coverage of scale and specialization trends.

The proprietary GPT-4o and GPT-4.1 serve as frontier references and, like the open models, were evaluated on all four services. Because their parameter counts and training data are undisclosed, we exclude them from the controlled scale and specialization analyses in RQ2 while retaining them for overall quality and cross-service comparisons (RQ1 and RQ3). GitHub Actions, the service used in prior work and the validation setting for our study (Section~\ref{sec:rq1}), is also the most common CI service in public code, providing a favorable setting for closed models and an upper-bound reference when interpreting per-service results.

Open models were run locally through Ollama with default 4-bit quantization, allowing even $32$B models to run on a single $80$\,GB GPU, and with their maximum supported context length. GPT models were accessed through the OpenAI API, and the exact Ollama model tags are listed in the replication package. All models used \texttt{temperature}=0; with fixed quantization and greedy decoding, generation is deterministic and reproducible across runs.

\subsection{Prompting and Generation}
We use a single standardized zero-shot prompt (Listing~\ref{lst:prompt}) to assess each model’s ability to turn a description into a structurally valid configuration without in-context examples, reflecting realistic first-attempt developer usage. The service-parameterized prompt requests a minimal, single, valid configuration with no comments or markdown. We stripped markdown fences from generated text before storage. Prompting all $16$ models across four services yielded $53{,}808$ configurations.

\begin{figure}[t]
\begin{lstlisting}[style=PromptStyle, label={lst:prompt}, caption={The standardized zero-shot prompt. \texttt{\{Service\}} is instantiated per CI service and \texttt{\{Description\}} holds the natural-language task.}]
@u@System:@ @s@You are a highly skilled software developer and
DevOps engineer.@

@u@User:@ @s@Given the description below, generate a minimal and
single valid {Service} YAML configuration. Do not provide
any reasoning. Only output the single YAML configuration,
with no comments, markdown, or extra formatting. The YAML
must match the described task exactly. Do not add any
unneeded directives unless told.@

@d@{Description}@
\end{lstlisting}
\vspace{-19pt}
\end{figure}

\subsection{Evaluation Metrics}
We score each generated configuration against its reference along two dimensions.

\smallskip\noindent\textit{\textbf{Similarity.}} We use eight metrics capturing lexical, semantic, and structural agreement: \emph{Cosine similarity} of sentence embeddings from \textit{all-MiniLM-L6-v2}~\cite{reimers2019sentence} (semantic metric); \emph{Euclidean similarity} (inverse embedding distance); \emph{BLEU}~\cite{papineni2002bleu}; \emph{CrystalBLEU}~\cite{eghbali2022crystalbleu}, which reduces the influence of shared $n$-grams; \emph{ROUGE-L}~\cite{lin2004rouge}; \emph{chrF}~\cite{popovic2015chrf}; \emph{tree edit similarity} on parsed YAML trees~\cite{zhang1989simple}; and \emph{canonicalized Levenshtein similarity}. All are scaled to $[0,1]$ (higher is better), except for cosine similarity, whose range is $[-1,1]$. A configuration is an \emph{exact} match when its canonicalized Levenshtein similarity to the reference is $1$. We use \textit{all-MiniLM-L6-v2} for the primary cosine metric because it is a compact encoder with strong semantic textual similarity performance relative to its size~\cite{reimers2019sentence,muennighoff2023mteb}, is widely used in recent software engineering studies~\cite{zhang2025little,cordeiro2024empirical}, and efficiently scores all $53{,}808$ outputs on CPU.

\smallskip\noindent\textit{\textbf{Validity.}} Similarity alone does not guarantee that a configuration is valid. Hence, we add two validity checks that range from lenient to strict. \emph{Parse validity} holds when the output loads as a YAML mapping. \emph{Schema validity} holds when a configuration also conforms to its service's JSON schema: the GitHub Actions~\cite{schemastore_github_actions} and Travis CI~\cite{schemastore_travis} schemas from SchemaStore, the CircleCI schema from the official CircleCI YAML language server~\cite{circleci_schema}, and the GitLab CI schema from the GitLab repository, used by its web editor~\cite{gitlab_schema}. We validate each output with the JSON Schema validator for the draft declared by its schema. We parse with a YAML 1.2 loader so that GitHub Actions \texttt{on:} is treated as a string rather than a boolean, and we use the same procedure for all four services to keep checks uniform and reproducible offline.
Each service also has a dedicated validator, namely \texttt{actionlint}~\cite{actionlint} for GitHub Actions, \texttt{circleci config validate}~\cite{circleci_validate} for CircleCI, the CI~Lint tool~\cite{gitlab_cilint} for GitLab CI, and \texttt{travis-yml}~\cite{travis_yml} for Travis CI. We do not use these as quality metrics as each assumes a complete, standalone configuration file, but our prompt requests a minimal configuration, and much of the documentation and generated outputs are intentionally partial fragments (Section~\ref{sec:data}). Linting these fragments would confuse incompleteness with incorrectness, and each validator applies service-specific, non-comparable rules. Schema conformance, applied uniformly across services, already captures the structural validity these validators check, which is why we use it as our strict check.

\subsection{Sampling and Open Coding}
\label{sec:coding}
To understand why models fall short (RQ4) and diverge on identical inputs (RQ5), we complement automatic metrics with manual qualitative analysis. Automatic scores measure distance between an output and its reference but do not reveal failure modes or causes of divergence, which coding uncovers under the same protocol. We draw stratified random samples of generated configurations with a target confidence level and margin of error, and analyze them using open coding~\cite{strauss1990basics}. For each sampled configuration, the primary coder (software engineering researcher with CI expertise) inspected three artifacts side by side: the task specification (service name and description), the documentation reference, and the model output. Coding followed constant comparison, starting from a small seed codebook and iteratively adding or merging codes as new discrepancy types emerged. Earlier items were revisited when the codebook changed, after which the finalized codebook was applied to the full sample and open codes were consolidated into higher-level categories.

\smallskip\noindent\textit{\textbf{Reliability.}} For independent coding, we assessed reliability through a second coder, a researcher with the same software engineering and CI expertise, on a service-stratified subsample using the finalized codebook and target service labels. We report Cohen's $\kappa$~\cite{cohen1960coefficient,landis1977measurement} between coders on both the three-point verdict (usable, partial, unusable) and the binary usable-versus-not decision. All disagreements were resolved through discussion to reach consensus (Section~\ref{sec:rq4}), and the subsample size and resulting coefficients are reported with the analyses they validate.

\subsection{Statistical Analysis}
\label{sec:statistical}
We use the Friedman test~\cite{friedman1937use} to compare models on per-description paired scores within each service, and to compare CI services on the $14$ common models' per-service means, following standard non-parametric practice~\cite{demvsar2006statistical,arcuri2011practical}. We use the Wilcoxon signed-rank test~\cite{wilcoxon1945individual} for paired contrasts, the Mann-Whitney U test~\cite{mann1947test} for the unpaired general-versus-code comparison, and apply Holm correction~\cite{holm1979simple} for multiple comparisons. We report effect sizes alongside $p$-values: Spearman's $\rho$~\cite{spearman1987proof} for model size–quality association, Cliff's $\delta$ for the general-versus-code contrast, and bootstrap $95\%$ confidence intervals for per-service means.

\section{Evaluation Results}
\label{sec:results}

\subsection{\textit{\textbf{RQ1: How well do LLMs generate CI configurations?}}}
\label{sec:rq1}
\subsubsection{\textbf{Motivation}}
To understand how well LLMs generate usable CI configurations, we establish a baseline at scale. Developers care about two distinct properties that a single score conflates: whether a configuration matches their intent and whether it is usable, meaning that it parses and conforms to the service schema. Prior work evaluated models on a single service using similarity alone, leaving unclear whether similar outputs are runnable. We evaluate all $16$ models across four services while separating similarity from validity, distinguishing lexically close but unusable outputs from usable but differently phrased ones.

\subsubsection{\textbf{Methodology}}
For each generated configuration, we compute the eight similarity metrics, using mean cosine similarity as the primary metric and releasing the others in the replication package. Cosine similarity is our main metric because it captures semantic agreement despite variation in CI YAML (reordered keys, quoting, whitespace), which lexical and edit-based metrics penalize even for equivalent configurations. We show below that all eight metrics rank models consistently, so emphasizing cosine does not affect our conclusions. We also compute exact-match rate (canonicalized Levenshtein = $1$) and two validity checks: YAML-mapping parse validity and conformance to the service JSON schema. We report per-model means across services and per-service validity. We compare our GitHub Actions results with the earlier GPT-4o evaluation~\cite{ghaleb2025canllmsci} to verify consistency and apply the statistical tests described in Section~\ref{sec:statistical} for model comparisons.

\medskip\subsubsection{\textbf{Results}}

Our GitHub Actions results are consistent with the earlier GPT-4o evaluation: GPT-4o achieves a mean cosine of $0.675$ and an exact-match rate of $2.6\%$, compared with the previously reported $0.69$ and $3\%$. Across the four services, the exact-match rate does not exceed $3.1\%$ (GPT-4o) and remains below $2\%$ for most models (Table~\ref{tab:rq1}), confirming at scale that LLMs rarely reproduce reference configurations. GPT-4.1 ($0.668$) and GPT-4o ($0.652$) lead on cosine \emph{similarity}, and the best open model (Qwen2.5-Coder-14B) achieves a relatively comparable score of $0.639$, while the weakest (CodeLlama-13B) achieves $0.532$. Models differ significantly within every service (Friedman, all $p<10^{-100}$). GPT-4.1 achieves a small but significant edge over GPT-4o (Wilcoxon $p<10^{-18}$, rank-biserial $r=0.19$), and the proprietary lead over the strongest open model is likewise significant but small (GPT-4.1 vs.\ Qwen2.5-Coder-14B, $p<10^{-39}$, $r=0.26$; the two GPT models vs.\ the $14$ open models, Cliff's $\delta=0.18$ per-configuration). The other seven metrics agree with cosine on model ranking: across the $16$ models, the per-model Spearman correlation between cosine and each other metric ranges from $0.73$ for CrystalBLEU to $0.92$ for chrF and Euclidean similarity, with all lexical, structural, and edit-based metrics in between. This indicates that the observed differences are driven by model outputs rather than by the choice of similarity metric, with CrystalBLEU as the only outlier because it downweights common $n$-grams shared by CI files.

\begin{table}[t]
\centering
\caption{Per-model quality across four services, grouped by size tier and ranked by mean cosine. \textbf{Bold} marks the best value within each tier/metric. \mgeneral{} denotes general and \mcode{} denotes code-specialized. Proprietary GPT models (shaded) are excluded from RQ2.}
\vspace{-5pt}
\label{tab:rq1}
\renewcommand{\arraystretch}{1.05}
\setlength{\tabcolsep}{5pt}
\footnotesize
\begin{tabular}{p{3.75cm} c r r r}
\toprule
\textbf{Model} & \textbf{Type} & \textbf{Cosine} & \textbf{Exact\%} & \textbf{Parse\%} \\
\midrule
\rowcolor[gray]{0.9} \multicolumn{5}{l}{\textit{\textbf{Proprietary (all services)}}} \\
\rowcolor[gray]{0.9} GPT-4.1 & \mgeneral{} & \textbf{0.668} & 2.7 & 96.7 \\
\rowcolor[gray]{0.9} GPT-4o  & \mgeneral{} & 0.652 & \textbf{3.1} & \textbf{98.3} \\
\midrule
\multicolumn{5}{l}{\textit{\textbf{Small (7 to 8B)}}} \\
Qwen2.5-7B       & \mgeneral{} & \textbf{0.609} & 1.1 & 97.2 \\
Llama3.1-8B      & \mgeneral{} & 0.602 & 1.1 & 96.6 \\
Qwen2.5-Coder-7B & \mcode{}    & 0.600 & 0.8 & 97.0 \\
CodeGemma-7B     & \mcode{}    & 0.571 & 0.9 & 96.4 \\
CodeLlama-7B     & \mcode{}    & 0.570 & \textbf{1.2} & \textbf{97.6} \\
\midrule
\multicolumn{5}{l}{\textit{\textbf{Medium (9 to 15B)}}} \\
Qwen2.5-Coder-14B & \mcode{}    & \textbf{0.639} & 1.5 & \textbf{99.0} \\
Qwen2.5-14B       & \mgeneral{} & 0.612 & 0.7 & 96.5 \\
Gemma2-9B         & \mgeneral{} & 0.586 & 1.5 & 94.0 \\
Gemma3-12B        & \mgeneral{} & 0.583 & \textbf{2.0} & 97.2 \\
CodeLlama-13B     & \mcode{}    & 0.532 & 0.7 & 96.3 \\
\midrule
\multicolumn{5}{l}{\textit{\textbf{Large ($>15$B)}}} \\
Qwen2.5-Coder-32B & \mcode{}    & \textbf{0.638} & 1.5 & 97.7 \\
Qwen2.5-32B       & \mgeneral{} & 0.636 & \textbf{1.9} & 97.5 \\
Gemma3-27B        & \mgeneral{} & 0.606 & 0.7 & 97.7 \\
CodeLlama-34B     & \mcode{}    & 0.588 & 1.3 & \textbf{98.6} \\
\bottomrule
\end{tabular}
\vspace{-5pt}
\end{table}

\begin{table}[t]
\centering
\caption{By service: mean cosine, parse validity, and schema validity (over all $16$ models).}
\vspace{-5pt}
\label{tab:validity}
\renewcommand{\arraystretch}{1.05}
\setlength{\tabcolsep}{6pt}
\begin{tabular}{p{4.15cm} r r r}
\toprule
\textbf{Service} & \textbf{Cosine} & \textbf{Parse\%} & \textbf{Schema\%} \\
\midrule
GitHub Actions & 0.641 & 97.1 & 66.6 \\
CircleCI & 0.550 & 94.9 & 56.3 \\
GitLab CI & 0.586 & 96.9 & 73.0 \\
Travis CI & 0.640 & 99.1 & 83.0 \\
\midrule
Overall & 0.606 & 97.1 & 70.8 \\
\bottomrule
\end{tabular}
\vspace{-10pt}
\end{table}

Configuration \emph{validity} reveals different trends for both YAML parsing and schema conformance (Table~\ref{tab:validity}). Almost all configurations parse as YAML mappings ($97.1\%$ overall), ranging from $94.0\%$ for Gemma2-9B to $99.0\%$ for Qwen2.5-Coder-14B, with slight variation across tiers ($96.9\%$ small, $96.6\%$ medium, $97.9\%$ large). In contrast, only $70.8\%$ conform to the service schema, with schema validity ranging from $56.3\%$ for CircleCI to $83.0\%$ for Travis CI. Schema validity is non-monotonic with scale ($67.7\%$ small, $63.2\%$ medium, $79.3\%$ large) due to a drop in Gemma models. Still, it differentiates models clearly: the best open models, Qwen2.5-32B ($82.2\%$) and Qwen2.5-Coder-32B ($83.0\%$), approach GPT-4.1 ($87.8\%$), while Gemma2-9B falls to $43.6\%$.

\begin{figure}[t]
\centering
\includegraphics[width=.9\linewidth]{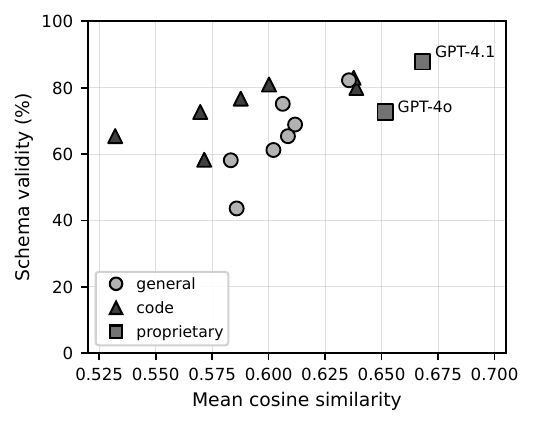}
\vspace{-10pt}
\caption{Similarity is not validity: each model's mean cosine (x) versus schema validity (y), across all four services and nearly uncorrelated across models.}
\vspace{-1pt}
\label{fig:rq1scatter}
\end{figure}

One illustrative case shows a mismatch between similarity and validity within a single service. On GitHub Actions, GPT-4o has the highest similarity of any model yet is only $17.2\%$ schema-valid, because it follows the minimal instruction literally and generates fragments in $80\%$ of cases (e.g., a bare \texttt{permissions} block) that match many fragment references but are not valid standalone files. GPT-4.1 instead generates complete GitHub Actions workflows ($2\%$ fragments) and achieves $94.8\%$ schema validity, indicating a sharp within-service split (McNemar $p<10^{-100}$). This completeness effect is specific to GitHub Actions: on CircleCI, GitLab CI, and Travis CI, GPT-4o produces complete files (near $0\%$ fragments) with $84\%$--$92\%$ schema validity, and its low GitHub Actions validity therefore reflects output completeness rather than an inability to produce valid YAML. On CircleCI, the pattern reverses: GPT-4o produces complete, schema-valid configurations that diverge from short references, ranking ninth of sixteen models on similarity and well below the best open model (rank-biserial $r=-0.28$). GPT-4.1, by contrast, ranks first on CircleCI, GitLab CI, and Travis CI and second on GitHub Actions, without GPT-4o's completeness-driven variability. Similarity thus substantially overstates usefulness: a lexically close configuration can still be unparseable, schema-invalid, or only a valid fragment rather than a runnable file. Fig.~\ref{fig:rq1scatter} shows both measures across all models.

\begin{tcolorbox}[rqbox]
\textbf{RQ1 Summary.} Similarity is an unreliable proxy for usable CI configurations: only $70.8\%$ of outputs satisfy service schema despite $97.1\%$ parsing successfully. Exact reproduction is rare ($\leq 3.1\%$). GPT-4o highlights this gap by achieving the highest GitHub Actions similarity but only $17.2\%$ schema-valid outputs. Schema validation is therefore essential for evaluating LLM-generated CI.
\end{tcolorbox}

\subsection{\textit{\textbf{RQ2: How do model scale and code specialization correlate with quality?}}}
\label{sec:rq2}
\subsubsection{\textbf{Motivation}}
Model choice for CI generation involves a trade-off between cost and quality. Larger models require substantially more resources, while code-specialized models are often expected to perform better on structured targets such as YAML. Whether either factor improves CI configuration quality remains unclear: the constrained nature of CI files may limit scale benefits, and prior single-service evidence suggests code specialization may not help. We therefore isolate model scale and code specialization across services to determine which, if either, provides a measurable advantage.

\medskip\subsubsection{\textbf{Methodology}}
Across the $14$ open models, we measure the association between parameter count and quality using Spearman's $\rho$ on per-model means, with both semantic (mean cosine) and structural metrics (tree edit, ROUGE-L, chrF) to capture effects beyond embedding similarity. We compare size tiers on paired per-configuration scores using Mann-Whitney tests and Cliff's $\delta$. To evaluate code specialization, we first compare general and code-specialized models in aggregate, then perform the cleaner paired analysis: each general model versus its code-specialized counterpart on identical descriptions using the Wilcoxon signed-rank test. The three size-matched Qwen2.5 pairs ($7$B, $14$B, $32$B) isolate specialization by holding scale and model family fixed.

\medskip\subsubsection{\textbf{Results}}
Scale improves structural validity, but its effect on embedding similarity is modest within this lineup of mid-to-large models. Parameter count correlates significantly with the structural metrics (tree edit $\rho=0.65$, ROUGE-L $\rho=0.61$, chrF $\rho=0.65$, all $p<0.05$), yet its correlation with mean cosine is positive and not significant at this sample size ($\rho=0.47$, $p=0.09$, $n=14$). Mean cosine is essentially flat across the small and medium tiers (both $\approx0.59$) and only modestly higher in the large tier ($0.62$). Across the $47{,}082$ paired per-configuration scores from the $14$ open models, the small-to-medium contrast is not even significant (Cliff's $\delta=-0.010$, Mann-Whitney $p=0.12$), and although the medium-to-large and small-to-large contrasts are significant ($p<10^{-3}$) their effect is negligible in magnitude (Cliff's $\delta=-0.068$ and $-0.081$, both below the $0.147$ negligible threshold~\cite{cliff1993dominance,romano2006appropriate}). Scale therefore has at most a practically negligible effect on embedding similarity once every model is already at least $7$B parameters. The usable rate from the open coding (Section~\ref{sec:rq4}) provides a clearer view of the scale effect, increasing monotonically from $34.7\%$ for small models to $40.0\%$ for medium models and $46.9\%$ for large models, with the proprietary models highest at $50.0\%$.

\begin{figure*}[t]
\centering
\includegraphics[width=.9\textwidth]{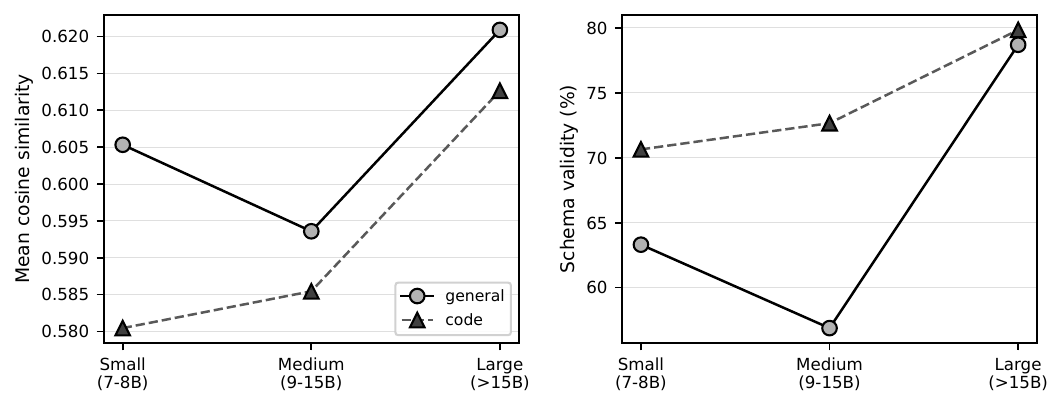}
\vspace{-10pt}
\caption{Quality by size tier for code-specialized and general open models (tier means): similarity (left) and schema validity (right). The two types converge only at the large tier.}
\vspace{-7pt}
\label{fig:rq2scale}
\end{figure*}

Code specialization gives no clear aggregate benefit on similarity, and its effect is non-monotone in scale. General and code models have almost identical mean cosine ($0.605$ versus $0.591$; Mann-Whitney $p=0.46$; Cliff's $\delta=0.27$, small), though code models are somewhat more schema-valid in aggregate ($73.8\%$ versus $65.0\%$). The within-family contrasts, each measured on the identical descriptions evaluated across all four services, explain the flat aggregate effect: the code-specialized variant wins only at the two larger Qwen sizes (Qwen2.5-Coder-14B $0.639$ versus Qwen2.5-14B $0.612$, $p<10^{-27}$; Qwen2.5-Coder-32B $0.638$ versus Qwen2.5-32B $0.636$, $p<10^{-2}$), whereas the general model wins at the smallest Qwen size (Qwen2.5-7B $0.609$ versus Qwen2.5-Coder-7B $0.600$, $p<10^{-2}$) and in the Llama and Gemma families (Llama3.1-8B $0.602$ versus CodeLlama-7B $0.570$, $p<10^{-22}$; Gemma2-9B $0.586$ versus CodeGemma-7B $0.571$, $p<10^{-6}$). The Llama and Gemma pairs are not size-matched: each code-specialized model is smaller than its general counterpart ($7$B versus $8$--$9$B), and for Gemma a generation older. Part of the general model's advantage in these two families may therefore reflect scale rather than a cost of specialization, and only the Qwen pairs cleanly identify specialization effects. Even where significant, these within-family effects are small ($|r|\le0.22$) and point in opposite directions across families. Thus, code pretraining helps only at larger scale within a strong instruction-tuned family and is neutral or harmful otherwise. Fig.~\ref{fig:rq2scale} shows both patterns together: the general line declines at the medium tier before recovering, and the code and general lines converge only at the large tier on either factor.

\begin{tcolorbox}[rqbox]
\textbf{RQ2 Summary.} Scale predicts CI quality better than code specialization. The overall general-vs-code comparison shows no significant advantage (Cliff's $\delta=0.27$, small, $p=0.46$), and Qwen2.5 pairs indicate specialization helps only at $14$B and $32$B while slightly hurting at $7$B. Larger general models are thus strong baselines, with code specialization offering limited gains only at higher scales.
\end{tcolorbox}

\subsection{\textit{\textbf{RQ3: How does the target CI service affect quality?}}}
\label{sec:rq3}
\subsubsection{\textbf{Motivation}}
The four services in \dataset{} express the same automation concepts through different surface syntax, from GitHub Actions' \texttt{jobs} and \texttt{steps} to Travis CI's flat lifecycle phases, and they appear unevenly in public code and therefore in pretraining data. A model that writes fluent GitHub Actions may thus falter on CircleCI's orbs or GitLab's stages, and a study that measures only the most common service would produce an optimistic estimate of performance that may not transfer across services. Quantifying the effect of target service reveals where LLM assistance is reliable and where human oversight remains necessary, while directly testing whether single-service conclusions from prior work generalize across the CI ecosystem.

\medskip\subsubsection{\textbf{Methodology}}
Because all $16$ models were run on every service, the cross-service comparison uses the full set. For each model, we compute its mean cosine per service, then test whether the target service changes quality with a Friedman test on the paired per-model means, which controls for model identity by treating each model as its own block. A significant omnibus result is followed by Holm-corrected Wilcoxon signed-rank contrasts between service pairs. We report bootstrap $95\%$ confidence intervals on each service mean and quantify every pairwise gap with Cliff's $\delta$ on the per-configuration scores, so that significance at this sample size is not mistaken for a large effect.

\medskip\subsubsection{\textbf{Results}}
The target service significantly affects quality (Friedman $\chi^2=43.3$, $p<10^{-8}$). Mean cosine is highest for GitHub Actions ($0.641$, $95\%$ confidence interval $[0.630,0.651]$) and Travis CI ($0.640$, $[0.618,0.664]$), which are statistically tied, then GitLab CI ($0.586$, $[0.570,0.601]$), and lowest for CircleCI ($0.550$, $[0.526,0.571]$). Five of the six pairwise contrasts are significant after Holm correction; only GitHub Actions and Travis CI are statistically indistinguishable ($\text{Holm}=0.94$, Cliff's $\delta=0.004$), while CircleCI is significantly the hardest service. The per-configuration effect sizes confirm the ordering and put it in perspective: CircleCI's disadvantage relative to the easy services is small (Cliff's $\delta$ up to $0.23$). The target service, though the largest single factor we measure, therefore moves similarity only modestly, though by more than model scale or specialization do. Two forces explain the CircleCI gap, both visible in the open coding (Section~\ref{sec:rq4}): its documentation contributes many server-installation references (Helm \texttt{values.yaml}, Kubernetes manifests) that are not workflow configurations, and models frequently leak GitHub Actions or GitLab idioms into CircleCI output. GitHub Actions and Travis CI are the easiest, with GitHub Actions benefiting from its prevalence in training data and Travis CI from its short references that are simple to reproduce. The manual coding agrees on the ordering and quantifies it: usable rates are highest for Travis CI ($56.0\%$) and GitHub Actions ($53.0\%$) and far lower for CircleCI ($26.7\%$) and GitLab CI ($27.1\%$), consistent with CircleCI also being the least schema-valid service.

\begin{tcolorbox}[rqbox]
\textbf{RQ3 Summary.} CI service dominates generation outcomes: CircleCI is the most challenging, while GitHub Actions and Travis CI are easier due to service-specific formats and idioms. Similarity differences are modest, but usability varies substantially ($56\%$ usable on Travis CI vs. $27\%$ on CircleCI). Single-service evaluations therefore overstate general performance.
\end{tcolorbox}

\subsection{\textit{\textbf{RQ4: Why do LLMs fall short in generation?}}}
\label{sec:rq4}
\subsubsection{\textbf{Motivation}}
Similarity and validity scores locate a model on a scale but do not reveal what a wrong configuration got wrong: a low cosine can indicate a missing step, an invented key, a wholesale misreading of the task, or merely a shorter but correct output than the reference. Those causes call for entirely different remedies, from better prompting to schema tooling to more training data, and no automatic metric distinguishes among them. A grounded, human-coded account of failure is therefore what turns the quantitative difference into an actionable diagnosis, and it is the only way to separate genuine capability limits from artifacts of using documentation as the ground truth source. We build such an account by open-coding a stratified sample of outputs into a failure taxonomy that later research and tooling can target directly.

\medskip\subsubsection{\textbf{Methodology}}
Following the protocol of Section~\ref{sec:coding}, we drew a stratified random sample of $385$ generated configurations ($95\%$ confidence level, $\pm 5\%$ margin of error) and open-coded them. We stratified the sample across the four services ($83$ GitHub Actions, $90$ CircleCI, $96$ GitLab CI, $116$ Travis CI), proportional to each service's share of the deduplicated benchmark and, within each service, approximately evenly across all $16$ models under a fixed random seed, ensuring every model was coded on every service. Because the sample is balanced across models rather than weighted by output volume, the taxonomy characterizes the range of failure modes and how they vary with model and service, not their frequency in the generated outputs.

\smallskip\noindent\textit{\textbf{Coding instrument.}} For each configuration, the coder recorded a three-part structured judgment for the item's target service. The first part is an overall \emph{verdict} on a three-point scale, defined so the boundaries are clearly decidable rather than a matter of degree:

\begin{itemize}[leftmargin=12pt, itemsep=1pt, topsep=2pt]
  \item \emph{usable}: runs on the target service, or would after only trivial edits (filling an obvious placeholder, renaming an identifier, reformatting), and expresses the documented behavior; a different-but-valid realization counts as usable;
  \item \emph{partial}: implements the mechanism described in the task but needs a non-trivial fix, such as a missing required key or a wrong value on the key the task is about;
  \item \emph{unusable}: does not express the documented behavior, is schema-invalid for the target service, or is degenerate.
\end{itemize}

A \emph{degenerate} flag is recorded on top of the verdict for output that is empty, prose instead of YAML, a stub, or repetitive, so that degeneration is counted without requiring the coder to separate ``wrong'' from ``broken'' on the main scale. When the reference is not a comparable standalone configuration for the target service (a bare fragment, a non-workflow artifact such as a Helm \texttt{values.yaml}, or a snippet in another tool's language), the coder does not grade by matching it but on whether the output expresses the documented concept for the service, and records a reference-quality code. The second part is one or more \emph{issue codes} that name each discrepancy (e.g., \texttt{missing\_key}, \texttt{wrong\_service\_idiom}, \texttt{hallucinated\_step}) and the YAML key it touches. The third is an \emph{attribution} of each discrepancy to one of three sources: the \emph{model} (a genuine error given a clear description), the \emph{description} (vague, underspecified, misleading, or based on a fragment or non-workflow reference, making the output defensible), or \emph{acceptable} variation (a legitimate alternative, not a defect). This attribution separates genuine model failures from artifacts of documentation-derived references.

\smallskip\noindent\textit{\textbf{Reliability.}} A service-stratified subsample of $\relsub$ configurations ($95\%$ confidence level, $\pm10\%$ margin of error)
was independently re-coded by a second researcher using the same codebook, with the target service provided for each item. We report Cohen's $\kappa$ on the three-point verdict and on the binary usable-versus-not decision.
Agreement on the three-point verdict was substantial, $\kappa=\kappaverdict$, and rises further on the binary usable-versus-not decision on which our findings are mainly based, $\kappa=\kappabin$. The $11$ disagreements out of $77$ are all one level apart, either \emph{usable} versus \emph{partial} or \emph{partial} versus \emph{unusable}, and each turns on a single sub-judgment the codebook leaves open, namely whether a required edit is trivial, or whether an output that uses the right mechanism with an invalid key is repairable enough to keep. None is a gross disagreement about whether the configuration works, which the earlier five-point scale had produced by splitting the same judgment across four boundaries. We resolved each disagreement through a consensus discussion meeting, in which the two coders re-examined the specification, reference, and output together, discussed the competing verdicts, and settled on the single verdict used in the analysis. Considering that all disagreements were adjacent rather than gross, discussion converged in every case, and each resolution clarified how the borderline sub-judgment above should apply so that the same reading was used consistently across the full sample.
We open-coded all $385$ configurations, yielding $560$ discrete coded issues, and consolidated the codes into six categories. We report the overall verdict distribution, category frequencies, and how these vary by service, model size, and model type.

\medskip\subsubsection{\textbf{Results}}
Overall, $41.3\%$ of sampled configurations are usable, $31.7\%$ are partial, and $27.0\%$ are unusable; $2.3\%$ fall under the degenerate flag (empty, non-YAML, or repetitive). Table~\ref{tab:taxonomy} shows the six failure categories. \emph{Misinterpretation} is the most common ($36.1\%$ of issues), covering wrong values, misread tasks, and wrong triggers, followed by \emph{Omission} ($24.1\%$). The remaining issues divide almost evenly: \emph{Reference/data quality} ($13.4\%$), a category absent from prior single-service coding, captures cases where the documentation reference is itself a fragment, a more elaborate configuration than the description warrants, a non-workflow artifact, or a snippet in another tool's language, so that a correct output is penalized; \emph{Superfluous additions} ($13.0\%$); and \emph{Invalidity} ($12.9\%$, schema violations, non-YAML output, and cross-service idiom leakage, reserved for output that does not parse or violates the service schema, not valid YAML that is merely off-target). Attribution reveals the source of the observed failures: of the $560$ coded issues, $432$ ($77.1\%$) are genuine model failures, $84$ ($15.0\%$) trace to the description or reference, and $44$ ($7.9\%$) are acceptable variation. Table~\ref{tab:examples} grounds the categories in concrete cases drawn from the coding.

\begin{table*}[t]
\centering
\caption{Representative examples of open-coded failure categories (model outputs are excerpted).}
\vspace{-5pt}
\label{tab:examples}
\renewcommand{\arraystretch}{1.15}
\resizebox{\linewidth}{!}{
\setlength{\tabcolsep}{5pt}
\begin{tabular}{@{}>{\RaggedRight\arraybackslash}p{3.3cm} >{\RaggedRight\arraybackslash}p{4.9cm} >{\RaggedRight\arraybackslash}p{5.4cm} >{\RaggedRight\arraybackslash}p{5.53cm}@{}}
\toprule
\textbf{Failure (model, service)} & \textbf{Task} & \textbf{Reference} & \textbf{Model output (excerpt)} \\
\midrule
Misinterpretation\newline (CodeLlama-13B, GitHub Actions) &
Run a job in a container with a Redis \emph{service} container. &
{\ttfamily services:\newline ~~redis:\newline ~~~~image: redis} &
{\ttfamily container: redis\newline steps:\newline ~~- run: npm test}\newline \emph{redis set as the job container, not a service} \\
\midrule
Invalidity: idiom leakage\newline (CodeGemma-7B, CircleCI) &
Specify the CircleCI configuration \texttt{version}. &
{\ttfamily version: 2.1} &
{\ttfamily jobs:\newline ~~build\_and\_test:\newline ~~~~runs-on: ubuntu-latest\newline ~~~~steps:\newline ~~~~~~- uses: actions/checkout}\newline \emph{GitHub Actions syntax in a CircleCI file} \\
\midrule
Omission\newline (Llama3.1-8B, GitLab CI) &
Add the Cobertura coverage report to \texttt{artifacts}. &
{\ttfamily artifacts:\newline ~~reports:\newline ~~~~coverage\_report:\newline ~~~~~~coverage\_format: cobertura} &
{\ttfamily artifacts:\newline ~~paths: [\$ARTIFACTS]}\newline \emph{omits \texttt{reports:coverage\_report}} \\
\midrule
Reference/data quality\newline (GPT-4o, GitHub Actions) &
Give the \texttt{--config-file-path} file that targets converted reusable workflows. &
{\ttfamily reusable\_workflows:\newline ~~- name: ...\newline ~~~~target\_url: ...} &
{\ttfamily jobs:\newline ~~audit: ...\newline ~~migrate: ...}\newline \emph{a workflow; the reference is a configuration file} \\
\bottomrule
\end{tabular}
}
\vspace{-7pt}
\end{table*}

The taxonomy varies across dimensions. By service, CircleCI has the highest share of reference-quality issues ($20.1\%$ of its issues, tied with GitHub Actions), reflecting the \texttt{values.yaml} and Kubernetes-manifest references of RQ3, alongside high invalidity ($17.5\%$ of its issues) from cross-service idiom leakage. GitLab CI failures are instead dominated by content errors split between misinterpretation ($36.5\%$) and omission ($33.3\%$), where models generate a plausible generic workflow that ignores the specific keyword the task requires (e.g., a generic \texttt{cache} block that drops the \texttt{cache:key:files} directive, or a plain \texttt{needs} list where the task requires \texttt{needs:project}). By scale, the \emph{nature} of failure shifts: invalidity peaks in the medium tier ($19.6\%$ of issues, against $12.8\%$ small) and then drops to $7.7\%$ in the large tier, while misinterpretation and superfluous additions rise (superfluous additions reach $14.6\%$ in the large tier). Larger models thus fail more by writing valid YAML that means the wrong thing or adds unrequested content than by generating broken output. Outright degeneration is concentrated rather than spread across scale: it accounts for only $2.3\%$ of verdicts overall and is essentially confined to CodeLlama, whose prose lead-ins drive the medium tier's $6.7\%$ degenerate rate and raise the code models' degenerate rate ($5.2\%$) well above the general models' ($0.0\%$).

\begin{table}[t]
\centering
\caption{RQ4 failure taxonomy over $560$ issues from $385$ configurations.}
\vspace{-5pt}
\label{tab:taxonomy}
\renewcommand{\arraystretch}{1.1}
\setlength{\tabcolsep}{4pt}
\begin{tabular}{p{2.8cm} r p{4.6cm}}
\toprule
\textbf{Category} & \textbf{\%} & \textbf{Representative codes} \\
\midrule
Misinterpretation & 36.1 & wrong value, misread task, wrong trigger \\
Omission & 24.1 & missing key/step, incomplete step \\
Reference/data quality & 13.4 & fragment, over-elaborate, non-workflow or wrong-language reference \\
Superfluous additions & 13.0 & hallucinated step, unnecessary configuration, over-specification \\
Invalidity & 12.9 & schema violation, non-YAML, idiom leakage \\
Surface variation & 0.5 & renamed identifier, reformatting \\
\bottomrule
\end{tabular}
\vspace{-7pt}
\end{table}

\begin{tcolorbox}[rqbox]
\textbf{RQ4 Summary.} Only $41.3\%$ of sampled outputs are usable as-is. Most failures are semantic rather than syntactic, driven by keyword misinterpretation ($36\%$) and omission ($24\%$), while reference artifacts ($13\%$) reveal a limitation of documentation-derived evaluation. As models scale, the remaining challenge shifts from producing valid YAML to producing the intended configuration.
\end{tcolorbox}

\subsection{\textit{\textbf{RQ5: Why do LLMs disagree in their generations?}}}
\label{sec:rq5}
\subsubsection{\textbf{Motivation}}
Results of RQ1 to RQ3 reveal that one model may score near-perfectly while others score near zero on the same input. Such disagreements could mean several different things, each with a different practical response. If they reflect genuinely different but defensible task interpretations, ensembling or sampling several models is the right lever; if they reflect a capability threshold that only some models clear, model selection is; and if they are an artifact of how the reference is written or scored, the implication concerns the benchmark design and evaluation procedure rather than the models. RQ4 characterizes failures across all cases, but the key disagreements are where the metric is least stable and therefore where its interpretation most needs investigation. This RQ addresses what drives the largest inter-model gaps.

\smallskip\subsubsection{\textbf{Methodology}}
To study why models diverge on identical inputs, we identify $852$ high-dispersion descriptions where at least $14$ models generated outputs, the best cosine score was $\geq0.72$, and the worst was $\leq0.35$. From these, we draw a service-proportional random sample of $86$ cases ($95\%$ confidence level, $\pm10\%$ margin of error, with median cosine range of $=0.79$). For each case, through collaborative review, we inspect the highest-scoring output and the failures of the two lowest-scoring outputs, then apply the constant-comparison open-coding procedure from Section~\ref{sec:coding} to assign the dominant driver of the best-versus-worst split. The resulting five categories are: \emph{completeness mismatch} (fragment references matched by one model but expanded into full workflows or under-produced by others), \emph{degenerate or invalid} (placeholder, prose, non-YAML, or hallucinated outputs), \emph{reference quality} (the reference is not a CI workflow configuration, e.g., a Helm \texttt{values.yaml}), \emph{valid alternative} (both outputs are valid but differ in form), and \emph{spec underdetermined} (the description admits multiple valid configurations). Table~\ref{tab:disagreement} summarizes the categories with representative examples.

\smallskip\subsubsection{\textbf{Results}}
Most inter-model disagreements do not reflect differences in generation ability (Table~\ref{tab:disagreement}). Completeness mismatch accounts for $51$ of the $86$ cases ($59\%$): the reference is a minimal fragment (e.g., \texttt{permissions:}, \texttt{dist:}, or \texttt{git:\,depth:\,false}), the winner reproduces that snippet, and lower-scoring models generate a complete, often valid, workflow around it. Another $15$ cases ($17\%$) are reference-quality artifacts, and $2$ ($2\%$) are valid alternatives or underdetermined specifications. Together, these account for $68$ of $86$ disagreements ($79\%$), indicating benchmark or output-completeness effects rather than capability gaps. Only $18$ cases ($21\%$) are genuine failures, where losing outputs are placeholder stubs, prose, or non-YAML.
The conclusion does not depend on ambiguous labels. Completeness mismatch alone accounts for $51$ of $86$ cases ($59\%$), a majority by itself, and is based on a structural distinction: a short reference, a winner matching that granularity, and losers expanding it into a full pipeline. The only ambiguous cases are weak CircleCI outputs labeled \emph{degenerate} versus \emph{reference-quality}, all within CircleCI's $30$ cases. Even if all were counted as genuine failures, the failure rate would rise only to at most $37\%$, so failures would still be a minority.

\begin{table*}[t]
  \centering
  \caption{The $86$ inter-model disagreements, grouped by dominant driver, with one representative case per category.}
  \vspace{-5pt}
  \label{tab:disagreement}
  \renewcommand{\arraystretch}{1.15}
  \setlength{\tabcolsep}{5pt}
  \resizebox{\linewidth}{!}{
  \begin{tabular}{@{}>{\RaggedRight\arraybackslash}p{2.5cm} r >{\RaggedRight\arraybackslash}p{6.9cm} >{\RaggedRight\arraybackslash}p{6.9cm}@{}}
  \toprule
  \textbf{Category} & \textbf{n (\%)} & \textbf{What drives the split} & \textbf{Representative case (service)} \\
  \midrule
  Completeness mismatch & 51 (59.3) &
  The reference is a minimal documented fragment; the winner generates exactly that snippet while the losers produce a complete, often valid, workflow (or under-produce a bare fragment against a full reference). &
  GitHub Actions's {\ttfamily permissions: \{\}}: the winner reproduces the fragment and scores near $1$, while the loser wraps it in a complete build workflow, valid YAML, and scores near $0$. \\
  \midrule
  Degenerate or invalid & 18 (20.9) &
  A losing model's output is genuinely broken: a placeholder stub, a prose essay, non-YAML, a hallucinated dump, or the wrong language. &
  Travis's \texttt{git:\,depth:\,false}: the winner reproduces it, whereas the loser returns {\ttfamily Here is a minimal\ldots} with no YAML body. \\
  \midrule
  Reference quality & 15 (17.4) &
  The gold reference is not a CI configuration, and matching it measures agreement with the reference rather than the ability to generate valid CI workflows. &
  CircleCI's reference is a Helm {\ttfamily values.yaml}: the model generates a YAML configuration instead. \\
  \midrule
  Valid alternative & 1 (1.2) &
  Best and worst are both valid configurations differing only in surface form or key spelling. &
  Both outputs are valid \texttt{variables:} blocks that differ only in content. \\
  \midrule
  Spec underdetermined & 1 (1.2) &
  The description admits several correct configurations, so different models pick different valid options. &
  ``Use Ubuntu 22.04'' maps to {\ttfamily dist: jammy} for one model and {\ttfamily os: ubuntu-22.04} for another model, and both are correct. \\
  \bottomrule
  \end{tabular}
  }
  \vspace{-5pt}
\end{table*}

The examples illustrate why similarity-based ranking can be misleading. On one GitHub Actions task whose reference is the single line \texttt{permissions: \{\}}, GPT-4o reproduces exactly that block and scores near $1$, whereas CodeLlama-7B and CodeLlama-34B wrap the same permission in a full build workflow and score near $0$ despite producing valid YAML. On a Travis CI task whose reference is \texttt{git:\,depth:\,false}, GPT-4o generates the bare key while other models generate complete language configurations. The ``best'' model on such fragments is the one whose default output completeness matches the reference. The proprietary models win most often (GPT-4.1 in $22$ cases and GPT-4o in $16$) because they tend to generate exactly the requested snippet, followed by mid-sized Qwen2.5-Coder-14B ($9$); the low scorer is typically CodeLlama, whose placeholder lead-ins (\texttt{Here is a minimal\ldots}) make CodeLlama-13B the worst output in $11$ of the $18$ genuine-failure cases and CodeLlama-7B in $10$.

The mechanism is also service-dependent. On CircleCI, disagreement splits between reference quality ($14$ of its $30$ sampled cases, where the reference is a CircleCI-server installation manifest such as a Helm \texttt{values.yaml} or Kubernetes \texttt{ConfigMap} rather than a \texttt{.circleci/config.yml} workflow) and genuine degeneration ($16$ of $30$, where a low-scoring model generates a placeholder stub or a prose explanation instead of YAML). In contrast, GitHub Actions, GitLab CI, and Travis CI disagreement is almost entirely completeness mismatch. Because high-dispersion references are often minimal fragments, the metric rewards models that generate the small requested snippet and penalizes models that produce complete (still valid) workflows. The result is a measure of output completeness rather than correctness, motivating the analysis of RQ6, where fine-tuning on these fragments improves similarity while reducing validity.

\subsection{\textit{\textbf{RQ6: To what extent can fine-tuning and schema-guided repair improve generation?}}}
\label{sec:rq6}
\subsubsection{\textbf{Motivation}}
Previous RQs evaluate models in their default form under zero-shot prompting and show that errors concentrate on misinterpreting or omitting task-specific keywords, with a smaller and largely mechanical invalidity tail rather than deep misunderstanding (RQ4). This suggests that targeted adaptation may reduce the gap without requiring larger models. We therefore evaluate two interventions: fine-tuning a modest open model on \dataset{} to learn service-specific conventions from description-to-configuration pairs, and a training-free schema-guided repair pass to measure how much invalidity can be corrected from validator feedback. Together, these interventions determine whether \dataset{} serves only as a diagnostic benchmark or also as a resource for improving configuration generation.

\medskip\subsubsection{\textbf{Methodology}}
We test two lightweight interventions against the validity and quality gap identified in RQ1. Both operate on the deduplicated \dataset{} pairs and reuse the exact zero-shot prompt from Listing~\ref{lst:prompt}, so that only the intervention, and not the prompt or decoding settings, changes. The first intervention is supervised fine-tuning, evaluated with $k$-fold cross-validation ($k=5$) stratified by service. We take one target configuration per unique description ($n=3{,}317$ unique descriptions; $46$ of the $3{,}363$ pairs share a description with a different reference snippet and collapse to a single target), ensuring that a description never appears in both training and test. Because \dataset{} is deduplicated (Section~\ref{sec:data}), no reference leaks across folds.

For each fold, we fine-tune on the remaining folds and generate on the held-out fold, then combine the held-out outputs across all folds so that every pair is evaluated out-of-fold and the fine-tuned and zero-shot models are compared on identical inputs. We fine-tune the small and large members of the strongest open family from RQ1, \ftbase{} and \ftbaseL{}, allowing the effect of adaptation to be measured at two scales rather than one. We use low-rank adaptation(LoRA)~\cite{hu2022lora} over $4$-bit quantized weights~\cite{dettmers2023qlora} (rank $16$, three epochs, cosine schedule), which keeps training within an academic GPU budget, and decode greedily to match the zero-shot runs. We refer to the resulting fine-tuned setting as \ftmodel{} (at both scales), giving later work building on \dataset{} a named similarity baseline for comparison, and share the generated outputs upon request. For each model, the baseline is the same base model generating zero-shot outputs for all pairs, regenerated with identical greedy settings to isolate the effect of adaptation.

\smallskip\subsubsection{\textbf{Results}}
\begin{figure}[ht]
\centering
\includegraphics[width=.9\linewidth]{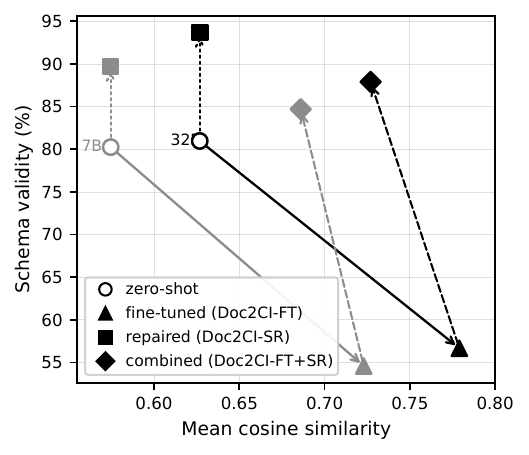}
\vspace{-12pt}
\caption{RQ6 interventions in the similarity-validity plane. Fine-tuning (\ftmodel{}) increases similarity while reducing validity, repair (\srmodel{}) increases validity at fixed similarity, and their combination (\ftsrmodel{}) improves over zero-shot on both axes.}
\vspace{-2pt}
\label{fig:rq6pareto}
\end{figure}

The two interventions shift different metrics in opposite directions, and that contrast is the result: fine-tuning greatly improves reference alignment but reduces standalone validity, while the repair pass boosts validity without affecting anything else.Fig.~\ref{fig:rq6pareto} places both moves in the similarity-validity plane: from the same zero-shot starting point, \ftmodel{} pulls down and to the right, while \srmodel{} pushes straight up.

Both bases are among the most schema-valid models in RQ1, with zero-shot schema validity already high (about $80\%$, well above the $71\%$ cross-model average), which makes them a demanding rather than a flattering starting point. On the combined out-of-fold test set, fine-tuning pulls each model's output strongly toward the documentation's own level of completeness (Table~\ref{tab:rq6}). For \ftbase{}, mean cosine rises from $0.575$ to $0.723$ and the exact-match rate rises from $0.5\%$ to $10.2\%$, about a twentyfold increase; for \ftbaseL{} the same shift appears, from $0.627$ to $0.779$ cosine and from $1.9\%$ to $17.3\%$ exact. Yet schema validity falls, from $80.3\%$ to $54.6\%$ for \ftbase{} (a change of $-25.7$ points, paired bootstrap $95\%$ confidence interval $[-27.6, -23.7]$) and from $81.0\%$ to $56.7\%$ for \ftbaseL{} ($-24.3$, $[-26.1, -22.5]$), while parse validity edges down from about $96\%$ to about $89\%$. Both movements are decisive and opposed. The similarity gain is a large paired effect (matched-pairs rank-biserial $r=+0.66$ for \ftbase{} and $+0.69$ for \ftbaseL{}, Wilcoxon $p<10^{-230}$), far larger than any model-scale or specialization effect in RQ2, whereas the validity loss is equally significant in the other direction: on the paired schema outcomes McNemar's test~\cite{mcnemar1947note} counts $1{,}072$ of \ftbase{}'s configurations turning from valid to invalid against only $221$ corrected ($p<10^{-130}$), and $995$ versus $189$ for \ftbaseL{}.

\begin{table}[t]
\centering
\caption{Fine-tuning and the combined \ftsrmodel{} pass (\srmodel{} repair over \ftmodel{}) versus greedy zero-shot on the combined out-of-fold \dataset{} test set ($n=3{,}317$). \textbf{Bold} marks the best under each measure within a base model.}
\label{tab:rq6}
\renewcommand{\arraystretch}{1.05}
\vspace{-5pt}
\setlength{\tabcolsep}{4pt}
\resizebox{\linewidth}{!}{
\begin{tabular}{p{3.7cm}rrrr}
\toprule
& \textbf{Cosine} & \textbf{Exact\%} & \textbf{Parse\%} & \textbf{Schema\%}\\
\midrule
\multicolumn{5}{l}{\textbf{\ftbase{}}}\\
~~~Zero-shot & 0.575 & 0.5 & 95.9 & 80.3\\
~~~\ftmodel{} & \textbf{0.723} & \textbf{10.2} & 89.0 & 54.6\\
~~~\ftsrmodel{} & 0.686 & 4.4 & \textbf{97.6} & \textbf{84.7}\\
\addlinespace
\multicolumn{5}{l}{\textbf{\ftbaseL{}}}\\
~~~Zero-shot & 0.627 & 1.9 & 95.5 & 81.0\\
~~~\ftmodel{} & \textbf{0.779} & \textbf{17.3} & 89.2 & 56.7\\
~~~\ftsrmodel{} & 0.727 & 8.1 & \textbf{98.6} & \textbf{87.9}\\
\bottomrule
\end{tabular}
}
\vspace{-7pt}
\end{table}

The drop is not noise but a direct consequence of what the model learns to imitate. As RQ5 showed, most \dataset{} references are minimal documentation fragments that illustrate a single keyword rather than a runnable file, and fine-tuning on them reproduces the same short, fragmentary output: the output matches the reference more closely (higher cosine and exact-match) but omits the wrapper keys a standalone file needs to satisfy the schema. The effect tracks how fragmentary each service's references are. It is most severe on GitHub Actions, where schema validity collapses from $70.2\%$ to $22.0\%$ for \ftbaseL{} because its references so often omit the surrounding \texttt{on:} and \texttt{jobs:} scaffold, and on CircleCI ($76.4\%$ to $28.1\%$), but is mild on GitLab CI and essentially absent on Travis CI ($91.4\%$ to $86.8\%$), whose short references are themselves usually schema-valid. Fine-tuning on \dataset{} therefore improves stylistic agreement with the documentation but should not be optimized on similarity alone, since on a fragment-heavy reference set that objective can come at the expense of the validity RQ1 found lacking.

The training-free repair pass moves the opposite way. Feeding the validator's own error back for a single correction raises schema validity from $81.0\%$ to $93.7\%$ for \ftbaseL{}, fixing $423$ of the $631$ invalid configurations ($67.0\%$) in one round, and from $80.3\%$ to $89.7\%$ for \ftbase{} ($312$ of $655$, $47.6\%$); both gains are significant by McNemar's test ($p<10^{-90}$). The pass is also monotone by construction: because it only revisits outputs the validator already rejected, it never turns a valid configuration invalid ($0$ regressions for either model), in sharp contrast to fine-tuning's wholesale validity loss. The recovered cases are concentrated exactly where fine-tuning did the most damage, namely GitHub Actions ($70.2\%$ to $93.9\%$ for \ftbaseL{}) and CircleCI, which confirms that much of the invalidity is mechanical: a missing or misplaced key that the schema message names precisely and the model can reinstate without any training and without disturbing similarity. The larger model repairs more of its own errors than the smaller one ($67\%$ versus $48\%$ of invalid outputs), and repair capability, like generation quality, therefore scales with model size.

\smallskip\noindent\textit{\textbf{Combining fine-tuning with repair.}} Because fine-tuning and repair move validity in opposite directions, we also apply them in sequence, running the \srmodel{} repair pass over the combined \ftmodel{} outputs, a combined setting we call \ftsrmodel{} (Table~\ref{tab:rq6}). This assesses whether one model can retain the fine-tuned similarity gains while recovering the standalone validity that fine-tuning gives up, since repair reinstates exactly the wrapper scaffold (\texttt{on:}, \texttt{jobs:}, \texttt{version:}) that fine-tuning learned to omit. For \ftbaseL{}, repair raises the fine-tuned model's schema validity from $56.7\%$ to $87.9\%$ and its parse validity from $89.2\%$ to $98.6\%$, while cosine settles back from $0.779$ to $0.727$ and exact-match from $17.3\%$ to $8.1\%$; \ftbase{} follows the same pattern, with schema validity increasing from $54.6\%$ to $84.7\%$ while cosine decreases from $0.723$ to $0.686$. The similarity gain largely survives, since \ftsrmodel{} retains $66\%$ of fine-tuning's cosine improvement over zero-shot at $32$B ($0.627$ to $0.727$) and $75\%$ at $7$B while ending well above the zero-shot schema rate ($87.9\%$ against $81.0\%$). Consequently, \ftsrmodel{} is the only setting that outperforms the zero-shot baseline on both similarity and validity simultaneously. It does not match the schema validity of repairing the zero-shot output directly ($93.7\%$), because the fragmentary text produced by fine-tuning is harder to complete than the complete zero-shot output.

The cosine drop from \ftmodel{} to \ftsrmodel{} is a large, highly consistent paired effect, since repair lowers cosine on nearly every configuration it edits (matched-pairs rank-biserial $r=-0.70$ for \ftbaseL{} and $-0.58$ for \ftbase{}, Wilcoxon $p<10^{-80}$), yet its absolute magnitude is modest: a mean cosine loss of $0.052$ (paired bootstrap $95\%$ confidence interval $[0.047, 0.057]$) for \ftbaseL{} and $0.037$ ($[0.033, 0.041]$) for \ftbase{}. The reduction is not an artifact of the mean, since the median and geometric-mean cosine follow the same trend (median $0.813$ to $0.755$ at $32$B and $0.748$ to $0.708$ at $7$B). On the paired schema outcomes, McNemar counts $1{,}037$ configurations turned valid against $0$ turned invalid for \ftbaseL{} ($1{,}000$ versus $0$ for \ftbase{}, both $p<10^{-300}$). The per-service pattern reflects the expected trade-off: on Travis CI, whose references are already valid, repair adds validity ($86.8\%$ to $95.6\%$) at almost no similarity cost ($0.818$ to $0.804$ cosine), whereas on the fragment-heavy GitHub Actions case it incurs a larger similarity cost ($0.836$ to $0.722$) to restore schema validity ($22.0\%$ to $80.3\%$).

\begin{tcolorbox}[rqbox]
\textbf{RQ6 Summary.} Fine-tuning on \dataset{} substantially improves reference similarity (cosine $0.627\!\rightarrow\!0.779$ at $32$B) but reduces standalone schema validity ($81.0\%\!\rightarrow\!56.7\%$) because the data contains many minimal fragments. Repair restores validity ($93.7\%$ schema validity), and combining the two achieves the best balance, improving both similarity and validity over zero-shot generation.
\end{tcolorbox}

\section{Discussion}
\label{sec:discussion}

\subsection{Implications}
\subsubsection{Similarity is necessary but not sufficient}
The gap between similarity and validity is the primary practical implication of our findings. A mean cosine near $0.7$ for the best models coexists with an exact-match rate of at most $3.1\%$ and, although parse failures are rare ($\sim$3\%), a schema-conformance rate of only $71\%$ across services. The GPT-4o case on GitHub Actions, where it achieves the highest similarity yet is among the least schema-valid because it generates fragments, makes the point vivid. Evaluations that report only similarity, as most prior work does, overstate how close LLMs are to usable CI automation. Validity checks against service schemas are cheap and should be standard in this setting.

\subsubsection{Output completeness over code specialization}
On a lineup of instruction-tuned models, the code-versus-general distinction is weak and non-monotone (RQ2): code specialization helps only at larger Qwen scale and is neutral or harmful elsewhere. What actually separates outputs is their \emph{completeness}, namely whether a model generates the configuration specified by the task or wraps it in a larger workflow, which is the dominant driver of cross-model disagreement (RQ5). The only residual output-discipline failures are CodeLlama's occasional prose lead-ins. For CI generation, teams should therefore prioritize scale and output discipline over the code-versus-general label.

\subsubsection{Documentation is an imperfect oracle}
Our references come from official documentation, which is authoritative but was not written to be a generation target. Many references are minimal fragments or non-workflow artifacts, which inflates apparent error under similarity and explains part of the low exact-match rate and the Travis CI divergence between similarity and manual usability. This is a caution for benchmark builders as much as for model developers, and it motivates references that are complete, structurally valid, and executable configurations.

\subsubsection{Repair, not fine-tuning, is the validity lever}
The two adaptations of RQ6 make the similarity-validity split concrete as an engineering choice. Fine-tuning on \dataset{}, released as \ftmodel{}, does exactly what supervised learning should, pulling outputs toward the training targets, which raises cosine and exact-match sharply at both model scales, and \ftmodel{} at $32$B achieves the highest similarity of any model in the study ($0.779$ cosine), which makes it the reference point later similarity-based work on this benchmark will compare against. Because those targets are largely minimal documentation fragments, however, the same move lowers standalone schema validity by about $25$ points, most of all on GitHub Actions, where references routinely omit the \texttt{on:} and \texttt{jobs:} scaffold. Optimizing a CI generator on reference similarity is therefore self-defeating for validity on a fragment-heavy reference set, a caution that generalizes to any benchmark developed from documentation snippets. The training-free \srmodel{} repair pass moves the other way, raising schema validity to $93.7\%$ and fixing two thirds of invalid configurations in a single round by feeding back the validator's own message, precisely because much of the invalidity is a mechanical missing or misplaced key rather than a conceptual error (RQ4). For teams, these results suggest prioritizing validation before fine-tuning: apply a validator-in-the-loop repair step before turning to fine-tuning when the goal is runnable output.

\medskip
\subsubsection{Practical recommendations}
~

\noindent\textbf{For CI engineers.} Treat LLM output as a draft, not a deliverable. With only $41.3\%$ of sampled configurations usable as-is and $27\%$ unusable, a generated file should be parsed and schema-checked before it is committed, and read against the specific keyword the task requires, since omission and misinterpretation of that keyword are the two most common failures. Practical value varies by service: expect the most help on Travis CI ($56.0\%$ usable) and GitHub Actions ($53.0\%$) and the least on CircleCI ($26.7\%$ usable, and only $56\%$ schema-valid across all models), and prefer a larger model, weighting output discipline over the code-versus-general label. Where resources permit, prefer larger models despite their higher latency and memory requirements, prioritizing scale and output discipline over the code-versus-general label.

\smallskip\noindent\textbf{For CI platforms.} Many LLM failures are detectable before execution. Editor-integrated schema validation, already available for human authors, would catch the small fraction of configurations that do not parse and the nearly one-third that violate the schema before they reach a workflow. Returning those validation errors to the model recovers two thirds of the invalid configurations in a single repair round (RQ6). Service-specific scaffolds could further reduce the cross-service idiom leakage that contributes to CircleCI's low scores.

\smallskip\noindent\textbf{For LLM developers.} On a capable lineup, outright degeneration is rare ($2.3\%$), with the remaining output-discipline failures largely limited to CodeLlama's occasional prose lead-ins. Constraining generation to valid YAML and the target service schema, whether through schema-aware decoding or the validator-in-the-loop repair of RQ6, removes most invalid outputs at low cost. Fine-tuning on documentation improves alignment with reference configurations but not standalone validity, making it complementary rather than a replacement. Because misinterpretation and omission are the dominant failure modes, retrieving service documentation during generation~\cite{lewis2020retrieval} to supply the required service-specific keywords is likely the higher-leverage intervention.

\smallskip\noindent\textbf{For researchers.} Evaluate across services and report validity alongside similarity. A single-service study would have missed that CircleCI is substantially harder than GitHub Actions, while a similarity-only evaluation would have missed that GPT-4o, the top model on GitHub Actions by similarity, is among the least schema-valid on that service. Documentation-derived benchmarks also require reference-quality controls: without them, the $13\%$ of coded issues (and $15\%$ by attribution) caused by fragment, non-workflow, or wrong-artifact references would be misinterpreted as model errors.

\subsection{Threats to Validity}
\label{sec:threats}

\subsubsection{Construct validity.}
We use each service's JSON schema, the same schemas used by IDE tooling, applied uniformly across all four services. Two schemas are maintained by the community-run SchemaStore project (GitHub Actions and Travis CI) and two by service-specific editor tooling (the CircleCI YAML language server and GitLab web-editor schema). Because schemas can lag behind newly introduced features, schema conformance approximates acceptance by the live service rather than certifying it; any lag makes our validity estimates conservative. We verified that all schemas resolve and validate known-good references but did not independently assess their feature coverage.
Although each service provides a dedicated validator (\texttt{actionlint}, \texttt{circleci config validate}, GitLab CI~Lint, and \texttt{travis-yml}), these tools assume complete standalone files and enforce service-specific rules. Applying them to our fragment-heavy benchmark would conflate incompleteness with incorrectness and prevent comparable cross-service measurement. We therefore report uniform schema conformance. Consequently, valid fragments matching fragment references may still be marked schema-invalid when a service schema requires a complete file; this distinction motivates the GPT-4o fragment analysis in RQ4.
We treat canonicalized Levenshtein similarity of $1$ as an exact match, allowing trivial formatting differences. However, fragment and non-workflow references can penalize correct outputs under similarity, which we quantify through the reference-quality category in RQ4. Finally, we use a single zero-shot prompt (Listing~\ref{lst:prompt}) for all $16$ models and four services. Although this may not elicit each model's best output, it avoids per-model prompt engineering and attributes quality differences to models and services rather than prompting. We leave more advanced prompting strategies to future work.

\subsubsection{Internal validity.}
Open coding involves human judgment, making coding reliability the main validity concern. Each discrepancy was attributed to model, description, or acceptable variation. For independent coding, a second researcher re-coded a stratified subsample using the three-point codebook, yielding substantial agreement on both the verdict ($\kappa=\kappaverdict$) and the usable-versus-not decision ($\kappa=\kappabin$). Disagreements were one level apart and resolved by consensus. RQ5 categories were instead assigned collaboratively: the two researchers reviewed each case together and agreed on its dominant driver, so no independent agreement coefficient is reported. This limitation has limited impact because the main failure category, \emph{degenerate or invalid}, is based on concrete failures (placeholder stubs, prose, non-YAML, or hallucinated outputs). Moreover, the key conclusion that most disagreements are not capability gaps follows from the completeness-mismatch majority, independent of the ambiguous CircleCI boundary (Section~\ref{sec:rq5}).
Residual boundary decisions (e.g., accepting a renamed job) could still shift category frequencies. The sample size of $385$, though grounded in a $95\%$/$\pm5\%$ design, is smaller than the full population.

The disagreement analysis (RQ5) draws a random $95\%$/$\pm10\%$ sample from a high-dispersion frame and therefore characterizes sharp-split cases rather than the benchmark as a whole. The fine-tuning of RQ6 uses a single fixed low-rank configuration (rank $16$, three epochs) rather than a swept hyperparameter search, and the magnitude of the validity drop could therefore shift with the training budget, though its direction is consistent across both model scales and follows from the fragment nature of the training targets.

\subsubsection{External validity.} We study four CI services, $16$ models, and English documentation. Other services, newer or larger proprietary models, few-shot or agentic prompting, and non-English descriptions may behave differently. The proprietary GPT models were evaluated on all four services alongside the open models. Possible exposure of the documentation during pretraining could inflate their scores, but the low exact-match rate (at most $3.1\%$) suggests limited leakage.

\subsubsection{Conclusion validity.} We use non-parametric tests appropriate for skewed similarity distributions~\cite{arcuri2011practical}, report effect sizes and confidence intervals, and apply Holm correction for multiple comparisons. Because $53{,}808$ configurations make even small differences statistically significant, we base substantive claims on effect sizes rather than $p$-values alone.

\section{Related Work}
\label{sec:related}
\subsection{CI configuration and its challenges} Hilton et al.~\cite{hilton2016usage} surveyed developers and found CI adoption limited by inexperience, while Widder et al.~\cite{widder2019conceptual} and Zampetti et al.~\cite{zampetti2020empirical} documented recurring configuration difficulties, and Vassallo et al.~\cite{vassallo2020configuration} detected configuration smells. Poor CI configurations were linked to long builds and failures~\cite{ghaleb2019duration,ghaleb2019noise,ghaleb2022interplay}, Ghaleb et al.~\cite{ghaleb2025cicd} characterized CI/CD configuration practices in open-source apps, and Rostami Mazrae et al.~\cite{rostami2023usage} analyzed the usage and migration of CI/CD tools. This body of work motivates automation but does not generate configurations.

\subsection{LLMs for CI and configuration generation}
Closest to our work, Mastropaolo et al.~\cite{mastropaolo2024toward} study the completion of partial GitHub Actions workflows, and Zhang et al.~\cite{zhang2024effectiveness} evaluate LLMs on manually crafted GitHub Actions scenarios, emphasizing syntactic validity and execution. Both target a single service and a small model set. Broader configuration-generation work translates descriptions into infrastructure or deployment files~\cite{rosa2023automatically,mehta2023automated}. A prior single-service evaluation~\cite{ghaleb2025canllmsci} examined six models on GitHub Actions only. Our study differs by spanning four CI services rather than one, covering $16$ models across scales and specializations rather than a handful, and combining similarity, validity, and a grounded failure taxonomy rather than reporting a single quality score. It further uses open coding to characterize failure modes and sources of disagreement across models.

\section{Conclusion}
\label{sec:conclusion}
We presented a large empirical study on generating CI configurations from natural language, across four CI services and $16$ LLMs, on a new benchmark called \dataset{} containing $3{,}363$ description-to-YAML pairs. LLMs remain far from reliable: exact reproduction rarely exceeds $3\%$, and although nearly all configurations parse, only $71\%$ conform to the official service schema. Larger models are more structurally faithful, though the gain in embedding similarity is not significant within our mid-to-large lineup, and code specialization confers no aggregate advantage, helping only at larger scale within a model family. The target service matters, with CircleCI hardest, and our failure taxonomy is led by misinterpretation and omission rather than invalidity, while cross-model disagreement reflects output completeness (the minimal documented fragment versus a complete workflow) rather than divergent understanding. These results argue for CI-aware, instruction-tuned models, for validity-based evaluation, and for editor-integrated schema support.

\medskip\noindent\textbf{Future work.} Our results suggest training CI-native models on large datasets of executable workflows and their run outcomes, allowing them to learn structural validity and service semantics directly rather than adapting from general code or text. Another direction is execution-grounded training that rewards configurations which parse, satisfy schema constraints, and execute successfully, making validation part of the optimization objective instead of a post-processing step. We also plan to explore few-shot, tool-augmented prompting, and agentic solutions in this context. Moreover, we aim to extend \dataset{} with executable references, more CI services, and newer proprietary models to enable consistent benchmarking.

\section*{Acknowledgment}
This work is supported by the Natural Sciences and Engineering Research Council of Canada (NSERC): RGPIN-2025-05897. The study was enabled in part by the Digital Research Alliance of Canada.

\balance
\bibliographystyle{IEEEtran}
\bibliography{refs}

\end{document}